\documentclass[%
 reprint,
 superscriptaddress,
 amsmath,amssymb,
aps,
]{revtex4-1}

\usepackage{graphicx}
\usepackage{dcolumn}
\usepackage{bm}
\usepackage{booktabs}
\usepackage{xcolor}
\usepackage[normalem]{ulem}

\newcommand{\wj}[6]{\left(
                           \begin{array}{ccc}
        \! #1\! & #2\!  & #3\!  \\
        \! #4\! & #5\!  & #6\!
                           \end{array}
                   \right)}

\begin{document}

\preprint{APS/123-QED}

\title{Enhancing bispectrum estimators for galaxy redshift surveys with velocities}
\author{Julius Wons}
\email{j.wons@unsw.edu.au}
\affiliation{Sydney Consortium for Particle Physics and Cosmology, School of Physics, The University
of New South Wales, Sydney NSW 2052, Australia}

\author{Emanuela Dimastrogiovanni}
\email{e.dimastrogiovanni@rug.nl}
\affiliation{Van Swinderen Institute for Particle Physics and Gravity, University of Groningen, Nijenborgh 4, 9747 AG Groningen, The Netherlands}\affiliation{School of Physics, The University of New South Wales, Sydney NSW 2052, Australia}

\author{Matteo Fasiello}
\email{matteo.fasiello@csic.es}
\affiliation{Instituto de F\'{i}sica Te\'{o}rica UAM-CSIC, calle Nicolás Cabrera 13-15, Cantoblanco, 28049, Madrid, Spain}
\affiliation{Institute of Cosmology and Gravitation, University of Portsmouth, PO1 3FX, UK}

\author{Jan Hamann}
\email{jan.hamann@unsw.edu.au}
\affiliation{Sydney Consortium for Particle Physics and Cosmology, School of Physics, The University
of New South Wales, Sydney NSW 2052, Australia}

\author{Matthew C. Johnson}
\email{mjohnson@perimeterinstitute.ca}
\affiliation{Perimeter Institute for Theoretical Physics, 31 Caroline St N, Waterloo, ON N2L 2Y5, Canada}
\affiliation{Department of Physics and Astronomy, York University, Toronto, ON M3J 1P3, Canada}

\date{\today}

\begin{abstract}
We forecast the ability of bispectrum estimators to constrain primordial non-Gaussianity using future photometric galaxy redshift surveys. A full-sky survey with photometric redshift resolution of $\sigma_z/(1+z)=0.05$ in the redshift range $0.2<z<2$ can provide constraints $\sigma(f^\mathrm{local}_\mathrm{NL})=3.4$, $\sigma(f^\mathrm{equil}_\mathrm{NL})=15$, and $\sigma(f^\mathrm{orth}_\mathrm{NL})=17$ for the local, equilateral, and orthogonal shapes respectively, delivering constraints on primordial non-Gaussianities competitive to those from the cosmic microwave background. We generalize these results by
deriving a scaling relation for the constraints on the amplitude of primordial non-Gaussianity as a function of redshift error, depth, sky coverage, and nonlinear scale cutoff. Finally, we investigate the impact that photometric calibration errors on the largest scales will have on the constraining power of future experiments. We show that peculiar velocities reconstructed via kinetic Sunyaev Zeldovich tomography can be used to mitigate the impact of calibration errors on primordial non-Gaussianity constraints. 
\end{abstract}

\maketitle


\section{Introduction}

Measurements of the Cosmic Microwave Background (CMB) anisotropies~\cite{Planck:2018vyg,ACT:2020gnv} and large galaxy redshift surveys~\cite{DES:2021wwk,eBOSS:2020yzd} have firmly established the current six-parameter cosmological standard model, $\Lambda$CDM. Within $\Lambda$CDM, the primordial density fluctuations that underlie the CMB anisotropies and that seeded the large scale structure (LSS) of the Universe are assumed to be nearly Gaussian. The discovery of deviations from Gaussianity would yield important information about the history of the early Universe and the fundamental interactions at play, with immediate consequences for both cosmology and particle physics (see~\cite{Bartolo:2004if,Chen:2010xka,Fergusson:2010dm,Byrnes:2014pja,Renaux-Petel:2015bja} for reviews on non-Gaussianity).~This tantalizing prospect has driven a tremendous effort in observational cosmology to place constraints on the amplitude of primordial non-Gaussianity (PNG), primarily through the study of three-point correlation functions, e.g. the bispectrum in momentum space.

The bispectrum dependence on its three momenta, subject to momentum conservation, is described by its so-called shape function.
It is often convenient to employ an (incomplete) basis of three such shapes or templates (local, equilateral, and orthogonal, see~\cite{Babich:2004gb,Senatore:2009gt}) to capture the predictions associated with various models of inflation. One may also quantify the degree to which a given shape is well-described by a linear combination of the standard templates by defining a scalar product among bispectrum shapes~\cite{Creminelli:2010qf}. The tightest existing constraints on the amplitude of PNG come from the \text{Planck} mission~\cite{Planck:2019kim}, with statistical errors of $\sigma(f_{\rm NL}^{\rm local} ) \simeq 5$, $\sigma(f_{\rm NL}^{\rm equil} ) \simeq 47$, and $\sigma(f_{\rm NL}^{\rm ortho} ) \simeq 24$. CMB-S4 is expected to improve these constraints by roughly a factor of $2$~\cite{CMB-S4:2016ple}. A natural target for future constraints on PNG is that of a sensitivity $\sigma(f_\mathrm{NL}) \alt 1$. This is because a non-Gaussianity of order one or larger would be very suggestive of a multi-field (or multi-clock) realization of inflation~\cite{Lyth:2001nq, Maldacena:2002vr,Acquaviva:2002ud,Zaldarriaga:2003my,Creminelli:2003iq,Creminelli:2004yq,Rigopoulos:2005ae,Wands:2007bd,Cabass:2022ymb}. It is worth stressing that multi-field models, besides being eminently testable, are also the most likely scenarios from the top-down perspective~\cite{Baumann:2014nda}.

Given that \textit{Planck} has exhausted nearly all of the information from the primary CMB temperature anisotropies, in the near future progress on non-Gaussianity  will have to rely mostly upon measurements of the LSS, which fills the volume of the observable Universe between us and the CMB sky and as such is, in principle, far more constraining.

Current bounds on PNG from measurements of the galaxy bispectrum in spectroscopic surveys~\cite{Cabass:2022wjy,Cabass:2022ymb}, as well as those stemming from the scale-dependent bias~\cite{Dalal:2007cu, Matarrese:2008nc} in the quasar power spectrum~\cite{Mueller:2022dgf} and other tracers~\cite{McCarthy:2022agq}, are order(s) of magnitude away from the $\sigma(f_\mathrm{NL}) \sim 1$ goal. However, near-term spectroscopic redshift surveys promise to match~\cite{DESI:2016fyo} CMB constraints and perhaps achieve $\sigma(f_\mathrm{NL}) \sim 1$, at least for the local shape~\cite{Dore:2014cca}. Similar constraints for the other shapes may be more difficult~\cite{Cabass:2022epm} even with futuristic spectroscopic surveys such as MegaMapper~\cite{Schlegel:2022vrv}. 

Spectroscopic surveys have the advantage of retaining truly three-dimensional information about the density field, but present the challenge of relatively low number densities due to limited observation time. Photometric surveys can be complementary, providing large number densities at the cost of missing much of the information about the density field along the line of sight. Forecasts indicate that photometric redshift surveys such as LSST can produce statistical error bars on $f_{\rm NL}$ that improve on existing constraints, and may be comparable to what is obtained with spectroscopic surveys~\cite{Karagiannis:2018jdt}.

Using the measured galaxy bispectrum to constrain PNG comes with several challenges. First, there are a number of modelling considerations for comparing measurements to theory, including relativistic light-cone and redshift-space effects (e.g.~\cite{Leicht:2020rct,DiDio:2018unb,Maartens:2020jzf}) and the modelling of nonlinear physics (see e.g.~\cite{Philcox:2022frc}). Another challenge, which will be the main focus of this paper, is large-angular scale systematic effects present in photometric surveys or spectroscopic surveys (such as DESI) whose targets are determined by imaging. These systematics include atmospheric blurring, unaccounted-for Galactic dust, and imperfect star-galaxy separation, among other effects~\cite{Huterer:2012zs,Muir:2016veb,DES:2015vnr}. The largest scales are most-affected because the larger the separation angle, the more time passes during the measurement and varying observational conditions lead to calibration errors. Excessive power has been found in several surveys~\cite{Ho:2012vy,Ho:2013lda,Pullen:2012rd,Agarwal:2013ajb,Agarwal:2013qta,Giannantonio:2013uqa} suggesting photometric calibration error, especially on the largest scales. As a result, information on these scales might not be accessible for the estimation of cosmological parameters and one may have to consider a large scale cutoff.

In this paper, we forecast the impact of this large scale cutoff on PNG constraints from future redshift surveys. We work in the light-cone basis and within a simplified model, considering only linear scales and ignoring redshift space distortions and relativistic light-cone effects. Note that the strongest contributions from redshift space distortions to the density field are on large angular scales~\cite{Challinor:2011bk} we discard. In the light-cone basis, we propose that a modified version of the KSW estimator~\cite{Komatsu:2003iq,Yadav:2007rk} can be applied to sets of redshift-binned galaxy maps as an optimal and unbiased estimator for the local, equilateral, and orthogonal bispectrum shapes. We present a simple scaling relation for the estimator variance in the light-cone basis and demonstrate that the estimator variance increases significantly for the local and orthogonal shapes when the largest angular scales are polluted by systematic effects. We then propose that there is an opportunity to restore much of the missing large scale information using the peculiar velocity field on large angular scales, reconstructed using the technique of kinetic Sunyaev Zel'dovich (kSZ) tomography~\cite{Zhang10d,Zhang:2015uta,Terrana:2016xvc,Deutsch:2017ybc,Smith:2018bpn,Cayuso:2021ljq}. Because kSZ tomography reconstructs velocities on large scales from small-scale statistical anisotropies, we do not expect the large scale velocity field to suffer from the same systematic errors as the galaxy survey. We explore to what extent reconstructed velocities can mitigate information loss due to large angular scale systematic effects in galaxy surveys.

\section{Power Spectra and Bispectra on the Lightcone}

We consider redshift-binned angular maps of the density and velocity, which for the linear modes we consider in this paper, are described by the multipole moments
\begin{equation}\label{eq:multipole_exp}
    a^X_{\ell m} = \int \frac{d^3k}{(2\pi)^3} \mathcal{T}^X_\ell (k) \mathcal{R}_k Y^*_{\ell m}(\hat{k})\,.
\end{equation}
where $X \in \{\delta^\alpha,v^\alpha\}$ denote the density or velocity field in a redshift bin $\alpha$, $\mathcal{R}(k)$ is the comoving curvature perturbation, and $\mathcal{T}^X_\ell (k)$ are bin-averaged transfer functions defined by:
\begin{equation}
\mathcal{T}^X_\ell (k) = \frac{1}{\Delta r} \int_{r_\alpha^{\rm min}}^{r_\alpha^{\rm max}} d r \ \mathcal{T}^X_\ell (k,r)\,,
\end{equation}
where $r$ is comoving radial distance. The bin width is defined by the bin boundaries $\Delta r = r_\alpha^{\rm max} - r_\alpha^{\rm min}$ and 
\begin{align}
    \mathcal{T}^{\delta}_\ell(k, r)&=4\pi i^\ell D_\delta(k, r) j_\ell(k r) \,,\label{eq:transfer_d}\\
    \mathcal{T}^{v}_\ell(k, r)& = - 4\pi i^\ell D_v(r) \frac{\partial j_\ell(k r) }{\partial r} \label{eq:transfer_v}\,.
\end{align}
In the following, we assume that the dark matter density field can be related to the galaxy density field by a known linear bias, and therefore use the term ``density'' interchangeably for galaxy number density and dark matter density.

Using the definitions above, the angular power spectrum of two fields $\{ X_1, X_2 \}$ is given by
\begin{equation}\label{eq:covariance}
    C^{X_1 X_2}_{\ell}=4\pi \int d\ln(k) \mathcal{T}^{X_1}_{\ell}(k)\mathcal{T}^{X_2}_{\ell}(k)P_\mathcal{R}(k)
\end{equation}
where $P_\mathcal{R}(k)$ is the power spectrum of the comoving curvature perturbation. The angle-averaged bispectrum between three fields $\{ X_1, X_2, X_3 \}$ is defined as
\begin{equation}
    \begin{split}
    B^{X_1X_2X_3}_{\ell_1\ell_2\ell_3}&= f_{\rm NL}
     \sqrt{\frac{(2\ell_1+1)(2\ell_2+1)(2\ell_3+1)}{4\pi}}\\&\times
        \wj{\ell_1}{\ell_2}{\ell_3}{0}{0}{0}b^{X_1X_2X_3}_{\ell_1\ell_2\ell_3} (f_{\rm NL} = 1)\,,
    \end{split}
\end{equation}
where the reduced bispectrum $b^{X_1X_2X_3}_{\ell_1\ell_2\ell_3}$ can be simply defined for separable shapes, whose three-point function is 
\begin{equation}
    \langle \mathcal{R}\mathcal{R}\mathcal{R} \rangle \propto \sum_i f^{(i)}(k_1) g^{(i)}(k_2) h^{(i)}(k_3) + 5 \ {\rm perm.}
\end{equation}
The functions $f^{(i)}(k_1)$, $g^{(i)}(k_2)$, and $h^{(i)}(k_3)$ are determined by the shape, and we have
\begin{equation}\label{eq:bispec}
    \begin{split}
        b^{X_1X_2X_3}_{\ell_1\ell_2\ell_3} &=
        \frac{1}{6}\sum_{i=1}^{N_{\rm fac}} \int_0^\infty dr\, r^2\sum_{\ell_1,\ell_2,\ell_3}\Big{[} 
        \mathcal{K}^{X_1}_{\ell_1}[f^{(i)}](r) \\ &\times
        \mathcal{K}^{X_2}_{\ell_2}[g^{(i)}](r)\,
        \mathcal{K}^{X_3}_{\ell_3}[h^{(i)}](r) + 5\,\rm{perm.}\Big{]}\,,
    \end{split}
\end{equation}
where $N_{\rm fac}$ gives the number of functions required; $N_{\rm fac}=1$ for the local shape and $N_{\rm fac}=4$ for the orthogonal and equilateral shapes. Following the notation of~\cite{Duivenvoorden:2019ses}, the $\mathcal{K}$-functionals appearing in Eq.~\eqref{eq:bispec} map the shape functions from Fourier to harmonic space, e.g.
\begin{equation}
    \mathcal{K}^{X}_{\ell}[f^{(i)}](r)\equiv \frac{2}{\pi}\int_0^\infty dk\, k^2 f^{(i)}(k)\mathcal{T}^X_\ell(k)j_\ell(kr)\,.
\end{equation}
The transfer functions $\mathcal{T}_\ell^{X}$ are computed using \texttt{CAMB}~\cite{Lewis:1999bs}. We find that an accurate computation of the reduced bispectrum requires fine sampling in both radial comoving distance and momenta.

\section{Estimator}
A general and optimal estimator for the amplitude of non-Gaussianity is~\cite{Komatsu:2001rj,Liguori:2010hx}
\begin{equation}\label{eq:estimator_actual}
    \begin{split}
        \hat{f}_\mathrm{NL}=&\frac{1}{6 \sigma^2}\sum_{\{\ell_i,m_i, X_i\}}(B_1)_{m_1m_2m_3}^{ \ell_1 \ell_2 \ell_3,X_1 X_2 X_3} \\
        \times&\Big{\{}\left[(C^{-1}a)^{X_1}_{\ell_1m_1}(C^{-1}a)^{X_2}_{\ell_2m_2}(C^{-1}a)^{X_3}_{\ell_3m_3}\right] \\
        &- \left[(C^{-1})^{X_1X_2}_{\ell_1m_1\ell_2m_2}(C^{-1}a)^{X_3}_{\ell_3m_3} + \mathrm{cyclic}\right]\Big{\}}\,,
    \end{split}
\end{equation}
where $(C^{-1})^{X_1X_2}_{\ell_1m_1\ell_2m_2}$ is the inverse of the covariance matrix defined by $C^{X_1X_2}_{\ell_1m_1\ell_2m_2}=\langle X_{\ell m}X'_{\ell' m'}\rangle$. For statistically isotropic fields, we restore the power spectrum defined in Eq.~\eqref{eq:covariance} via $C^{XX'}_{\ell}=C^{XX'}_{\ell m\ell'm'}\delta_{\ell\ell'}\delta_{m-m'}$. With this assumption, the normalization $\sigma^2$ describing the variance of the estimator reads
\begin{equation}\label{eq:estimator_master}
    \begin{split}
    \frac{1}{\sigma^2}= &\sum_{\{X_i\}}\sum_{\ell_1, \ell_2, \ell_3}
    \left(B_1\right)^{X_1 X_2 X_3}_{ \ell_1 \ell_2 \ell_3}(B_1^{*} )^{X_4 X_5 X_6}_{\ell_1 \ell_2 \ell_3}       \\ &\times
     \Big[(C^{-1})_{\ell_1}^{X_1X_4}(C^{-1})_{\ell_2}^{X_2X_5} (C^{-1})_{\ell_3}^{X_3X_6}\Big]
     \,,
    \end{split}
\end{equation}
where $B^*$ is the complex conjugate of the bispectrum. In our forecasts below, we will be mainly interested in the estimator variance Eq.~(\ref{eq:estimator_master}). Where relevant, we assume that the impact of fractional sky coverage $f_{\rm sky}$ due to masking can be incorporated through $\sigma^2 \rightarrow \sigma^2/f_{\rm sky}$.

Multiple approaches to reduce the complexity of the estimator have been introduced~\cite{Fergusson:2009nv,Bucher:2015ura}. In this work, we will apply the KSW estimator~\cite{Komatsu:2003iq,Yadav:2007rk}, which reduces the scaling with the maximal multipole $\ell_\mathrm{\max}$ from $\mathcal{O}(\ell_\mathrm{max}^6)$ to $\mathcal{O}(\ell_\mathrm{max}^3)$. 

Applying the estimator to $N_\mathrm{bin}$ 2D maps, such as those obtained from photometric redshift surveys, leads to another computational challenge. The sum over the 6 different $X_i$ in Eq.~\eqref{eq:estimator_master} leads to a scaling of $\mathcal{O}(N_\mathrm{bin}^6)$ where $N_\mathrm{bin}$ is the number of bins. Future photometric surveys will be able to divide redshift space into $\mathcal{O}(10-100)$ bins~\cite{LSSTScience:2009jmu,Dore:2014cca}, implying that there is a steep computational penalty. Fortunately, with increasing numbers of redshift bins, and at sufficiently high-$\ell$, the bin-bin correlations become concentrated near the diagonal. As we demonstrate in more detail below, for the density modes nearly all of the estimator variance is captured even when bin-bin correlations are completely ignored, dropping the total computational cost of the estimator to $\mathcal{O}(N^3_\mathrm{bin} \times \ell_\mathrm{max}^3)$. When velocities are included, there are significant bin-bin correlations, but these are nevertheless of relatively compact support in bin-space, and the range in $\ell_\mathrm{max}$ we consider below is sufficiently small that it is still  computationally feasible to evaluate the estimator. For the near-term photometric redshift surveys such as LSST, with $N_\mathrm{bin} \sim 20$ and an $\ell_\mathrm{max}$ that includes all linear scales, we estimate this to be roughly $100 \times$ the computational cost for the similar analysis of \textit{Planck} data~\cite{Planck:2019kim}.
    
\section{Forecast Setup}\label{sec:setup}
Our first goal is to forecast the constraining power of LSS measurements for primordial non-Gaussianity, including limited redshift resolution (as a model for photometric galaxy redshift surveys), restricting to linear scales, and subject to large angular-scale systematics. In our forecast, we neglect redshift space distortions, magnification bias, and other contributions to the observed galaxy number counts. While including these effects will have some numerical impact on our forecast, we do not expect them to change our conclusions. For all our forecasts, we use the \textit{Planck} 2018 best-fit cosmology~\cite{Planck:2018vyg} and assume full sky coverage. 

We further assume a linear galaxy bias and that shot noise can be neglected in measurements of the galaxy density over the redshift range $0.2<z<2$; this is a reasonable assumption for future surveys such as LSST. We consider three fiducial redshift uncertainties, $\sigma_z/(1+z) = \{ 0.05, 0.02, 0.005\}$, which are used to map onto a number of redshift bins $N_{\rm bin} = \{20, 48, 185 \}$ over the redshift range we consider. The largest of these uncertainties could be obtained by near-term surveys such as LSST and SPHEREX, while the smallest uncertainty is chosen based on target parameter constraints. The smallest angular scales we consider, i.e. the maximal multipole $\ell_\mathrm{max}$, for the density field at a given redshift is set by the nonlinear scale $k_{\rm NL}(\bar{z})$ by
\begin{equation}\label{eq:nl_scaling}
\ell_\mathrm{max}(\bar{z}) \sim k_{\rm NL}(\bar{z}) r(\bar{z})\,,
\end{equation} 
where $\bar{z}$ is the mean redshift of the redshift bin and $r(\bar{z})$ its mean radial comoving distance. We define the nonlinear scale $k_{\rm NL}$ as the scale where the linear power spectrum deviates more than $1\%$ from the nonlinear one. We used the HMCode 2020~\cite{Mead:2020vgs} to calculate the nonlinear power spectrum. Note that $\ell_\mathrm{max}$ increases with redshift both due to the growth of structure and the projection.

In addition to the density field, we consider the peculiar velocity field as reconstructed using kinetic Sunyaev Zel'dovich (kSZ) tomography. This technique exploits the statistical anisotropy in the cross-correlation of the cosmic microwave background (CMB) temperature anisotropies and a galaxy survey to reconstruct the radial peculiar velocity field. The reconstruction has the highest fidelity on large angular scales, and therefore when velocities are considered below, we use $\ell_{\mathrm{max},v}=20$ to ensure that we are well within the signal-dominated regime~\cite{Deutsch:2017ybc}. 

\section{Results}\label{sec:results}
First, we investigated the dependency of the estimator variance on the redshift resolution, varying the number of bins between $1\leq N_\mathrm{bin} \leq128$ for the density field and found that the scaling is a power law in the number of bins
\begin{equation}\label{eq:bin_scaling}
\sigma(f_\mathrm{NL})\propto 1/N^{\gamma/2}_{\mathrm{bin}}\,,
\end{equation}
where the parameter $\gamma$ depends on $N_\mathrm{bin}$. If the bins are uncorrelated independent of their size, we expect $\gamma=1$ as we sum over $N_\mathrm{bin}$ maps in Eq.~\eqref{eq:estimator_master}. In reality, there are bin-bin correlations that encode radial density perturbations, which will be more important as the bin size is decreased. However, computing the forecasted estimator variance when including all bin-bin correlations we find that $\gamma = 1$ is a reasonable approximation for the redshift resolutions we consider -- the constraining power of near-term surveys is therefore mainly coming from angular correlations. Ignoring the correlation leads to an overestimate of the estimator variance between $5-15\%$ depending on the shape and redshift resolution. We neglect bin-bin correlations for the density field in the following; the correlated result gives strictly better constraints meaning the presented results are an upper bound.

\begin{table}[ht]
    \centering
    \begin{tabular}{c|c| c c c}
    \hline\hline
        $\sigma_z/1+z$ & $N_\mathrm{bin}$& ~ local ~ & \,equilateral\, & \,orthogonal\, \\ \hline
        0.05 & 20 & 3.4 & 15 & 17  \\
        0.02 & 48 & 2.1 & 8.5 & 10  \\
        0.005 & 185 & 1 & 4.3 & 5.2 \\
             \hline\hline
    \end{tabular}
    \caption{Results of the forecast of $\sigma(f_\mathrm{NL})$ for a full-sky density survey in the redshift range $0.2<z<2$. Results are presented for different numbers of bins $N_\mathrm{bin}$ determined by the redshift resolution $\sigma_z/1+z$.}
    \label{tab:results}
\end{table}

In Tab.~\ref{tab:results}, we present the $1\sigma$ constraints on $f_\mathrm{NL}$ using the density field only, in the uncorrelated bin approximation. The results for the lowest redshift resolutions (top two rows) are competitive with current and forecasted CMB constraints~\cite{Planck:2019kim,CMB-S4:2016ple}. To reach $\sigma(f_\mathrm{NL})\leq1$, the required resolution is roughly $\sigma/(1+z) = 0.005$. This implies that practically speaking a spectroscopic survey is required to reach $\sigma(f_\mathrm{NL})\leq1$ over the redshift range we consider. 

Another way to increase the sensitivity is by extending the redshift range to higher redshifts. At the lowest redshifts, the linear regime only spans up to $\ell_\mathrm{max}=50$, while the highest redshift bin reaches $\ell_\mathrm{max}>800$ -- there are many more linear modes at high redshift. The $\ell$-range could also be increased by an ansatz to describe fluctuations on nonlinear scales. Applying Eq.~\eqref{eq:nl_scaling}, the nonlinear scale varies in the range $0.05 - 0.15\,\mathrm{Mpc}^{-1}$ depending on the redshift. Assuming the model to describe the nonlinear scales that allows us to increase the nonlinear scale by a factor $\xi_\mathrm{NL}$, then the $\ell$-range would increase by the same factor.

Varying the parameters of the forecast, we obtain the scaling relation 
\begin{equation}\label{eq:sigma_scaling}
    \sigma(f_\mathrm{NL})\approx 
    A_0
    \sqrt{f_\mathrm{sky}}
    \left(\frac{0.05}{\frac{\sigma_z}{1+z}}\right)^{\frac{1}{2}}
    \left(\frac{r(z=2)}{r(z_\mathrm{max})}\right)^2 {\frac{1}{\xi_\mathrm{NL}}}
    \,,
\end{equation}
with $A_0$ being 3.4 for the local shape and $15$ and $17$ for the equilateral and orthogonal shapes respectively; the maximal redshift is expressed in terms of the comoving distance $r(z_\mathrm{max})$. The scaling relation only holds if shot noise can be neglected, which might no longer be the case if $z_\mathrm{max}$ or $\xi_\mathrm{NL}$ become too large.

\begin{figure}[ht]
    \centering
    \includegraphics[width=0.99 \columnwidth]{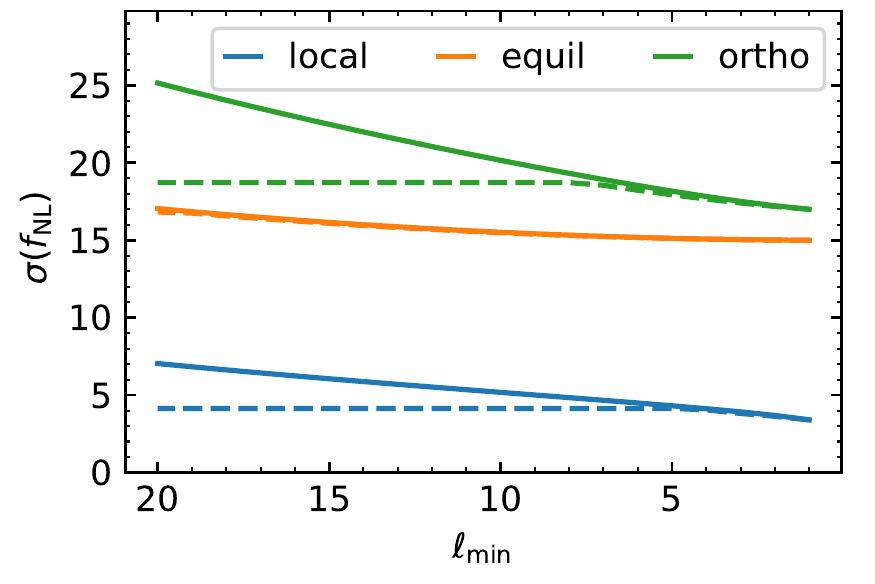}
    \caption{Results for $\sigma(f_\mathrm{NL})$ with truncated low multipole. In bold, $\sigma(f_\mathrm{NL}^\delta)$ of the density map in dependence on the $\ell_\mathrm{min}$. The dashed lines show $\sigma(f_\mathrm{NL}^{\delta+v})$ when including low-$\ell$ velocity multipoles to the bold lines. For both lines we used $N_\mathrm{bin}=20$.}
    \label{fig:lmin}
\end{figure}

Finally, we studied how photometric calibration errors and other large angular scale systematics~\cite{Huterer:2012zs} affected the forecast. We demonstrate this by truncating the fiducial model describe in Sec.~\ref{sec:setup} at the largest scales as shown in Fig.~\ref{fig:lmin}. The bold lines show the forecasted constraints on $f_{\rm NL}$ as a function of the lowest multipole $\ell_\mathrm{min}$, i.e. $\sigma(f^{\delta}_{\mathrm{NL},\ell_\mathrm{min}})$. For the local shape, the percentage deterioration is the largest because the dominant contribution comes from configurations combining large and small scales. The dominant contribution for the equilateral shape comes from combining equal scales, thus removing large scales does not impact the estimator as significantly as it does for the local shape.

The lost information can be partially restored by including velocities. The dashed lines in Fig.~\ref{fig:lmin} show the combined estimator $\sigma(f^{\delta+v}_{\mathrm{NL},\ell_\mathrm{min}})$ of the truncated density field together with the low-$\ell$ velocity field up to $\ell_\mathrm{max,v}=20$. When including velocities it is necessary to consider bin-bin correlation as large parts of the information come from the correlation. The scaling of the estimator remains the same albeit with a different amplitude. By comparing the truncated uncertainty $\sigma(f^{\delta}_{\mathrm{NL},\ell_\mathrm{min}})$ and the velocity only uncertainty $\sigma(f^{v}_{\mathrm{NL}})$ at $\ell_\mathrm{min}$, we can then accurately extrapolate the value for $\sigma(f^{\delta+v}_{\mathrm{NL},\ell_\mathrm{min}})$ at $\ell_{\mathrm{max},v}$, without computing the fully correlated estimator at large $\ell$. 

For the local and orthogonal shape, large parts of the lost information can be restored. The velocities are therefore able to reduce the information loss of large scale uncertainties to less than $15\%$ as long as $\ell_{\mathrm{max},v}\geq\ell_\mathrm{min}$.

\section{Conclusion}
In this paper, we studied the constraining power of photometric redshift surveys on primordial non-Gaussianity. In Tab.~\ref{tab:results}, we present the results of our analysis, which shows that upcoming photometric surveys will be able to perform measurements of non-Gaussianities that are competitive with upcoming CMB surveys~\cite{CMB-S4:2016ple}. For attainable photometric redshift errors, we showed that nearly all of the variance for the KSW estimator is accounted for by considering only angular correlations, rendering an analysis of near-term surveys computationally feasible. The expected constraints promise to be competitive with current and upcoming CMB constraints. However, reaching the target of $\sigma(f_\mathrm{NL}) < 1$ from the bispectrum using only the linear scales is likely not attainable with future photometric surveys. A less conservative treatment of the nonlinear scales provides a promising way to improve the constraints as the number of multipoles in Eq.~\eqref{eq:nl_scaling} scales linearly with the nonlinear scale leading to an inverse linear scaling for the estimator. The addition of these higher multipoles will, on the other hand, require a more careful treatment of shot noise.

Further, we investigated the impact of the large angular-scale systematics afflicting photometric redshift surveys. Such systematics significantly affect constraints on the local and orthogonal shapes. We demonstrated that, to a large part, the losses can be negated by including the reconstructed radial velocity field obtained by kSZ tomography. Velocity reconstruction is expected to work best on large angular scales, making velocity reconstruction an ideal complement to the error-plagued large scale density maps.

Our findings are summarised in the scaling relation in Eq.~\eqref{eq:sigma_scaling} which provides an estimate of the constraining power on primordial non-Gaussianities for upcoming photometric galaxy surveys. The scaling relation holds as long as noise can be neglected.

We have studied the uncertainty of the bispectrum estimator to forecast constraints on $f_\mathrm{NL}$. Estimation of non-Gaussianities of a given photometric survey requires evaluating Eq.~\eqref{eq:estimator_actual}. For the first generation photometric galaxy surveys such as DES~\cite{DES:2021wwk}, the constraining power is rather limited, $\sigma(f_\mathrm{NL})\approx\mathcal{O}(100)$, primarily due to low sky coverage and low photometric redshift resolution.

To summarise, in this work we showed the remarkable constraining capabilities that future photometric galaxy surveys have
on primordial non-Gaussianity. They can provide an independent measurement with a constraining power
comparable to that of CMB measurements.

\begin{acknowledgments}
Most of the numerical calculations for this work were performed on the computational cluster Katana, supported by Research Technology Services at UNSW Sydney~\cite{Katana}. MCJ is supported by the National Science and Engineering Research Council through a Discovery grant. This research was supported in part by Perimeter Institute for Theoretical Physics. Research at Perimeter Institute is supported by the Government of Canada through the Department of Innovation, Science and Economic Development Canada and by the Province of Ontario through the Ministry of Research, Innovation and Science. MF would like to acknowledge support from the “Ram\'{o}n y Cajal” grant RYC2021‐033786‐I, MF's work is partially supported by the
Agencia Estatal de Investigaci\'{o}n through the Grant IFT
Centro de Excelencia Severo Ochoa No CEX2020-001007-
S, funded by MCIN/AEI/10.13039/501100011033.
\end{acknowledgments}

\bibliography{apssamp}

\end{document}